\documentclass[%
 reprint,
 superscriptaddress,
 amsmath,amssymb,
 aps,
]{revtex4-2}

\usepackage{graphicx}
\usepackage{dcolumn}
\usepackage{bm}
\usepackage{subfigure}
\usepackage{xcolor}

\begin{document}

\preprint{APS/123-QED}

%
\title{Colossal Atomic Force Response in van der Waals Materials \\Arising From Electronic Correlations}

\author{Paul Hauseux}
\affiliation{Department of Engineering,
	University of Luxembourg, L-4365 Luxembourg
}%

\author{Alberto Ambrosetti}
\affiliation{ Dipartimento di Fisica e Astronomia,
	Universit\`a degli Studi di Padova, 35131 Padova, Italy
}%

\author{St\'ephane P. A. Bordas}
\affiliation{Department of Engineering,
	University of Luxembourg, L-4365 Luxembourg
}%

\author{Alexandre Tkatchenko}
\affiliation{Department of Physics and Materials Science,
University of Luxembourg, L-1511 Luxembourg
}%

\date{\today}

\begin{abstract}
Understanding static and dynamic phenomena in complex materials at different length scales requires reliably accounting for van der Waals (vdW) interactions, which stem from long-range electronic correlations. While the important role of many-body vdW interactions has been extensively documented when it comes to the stability of materials, much less is known about the coupling between vdW interactions and atomic forces. Here we analyze the Hessian force response matrix for a single and two vdW-coupled atomic chains to show that a many-body description of vdW interactions yields atomic force response magnitudes that exceed the expected pairwise decay by 3-5 orders of magnitude for a wide range of separations between the perturbed and the observed atom. Similar findings are confirmed for graphene and carbon nanotubes. This colossal force enhancement suggests implications for phonon spectra, free energies, interfacial adhesion, and collective dynamics in materials with many interacting atoms. 
\end{abstract}

\maketitle

Many phenomena in materials involve the interaction between electrons and phonons, and more generally between electrons and atomic lattices. Such interactions are the cornerstone of many-body physics in condensed matter and they contribute to fundamental quantum phenomena such as the temperature-dependence of the electrical conductivity in metals~\cite{giustino,ashcroft}, Cooper-pair formation in superconductivity~\cite{supercond,cooper}, thermalization and transport of charge carriers~\cite{hotcarr}, and magnetic properties of molecules and materials~\cite{ashcroft}. The dimensionality, system size, and nature of interatomic interactions (strong or weak bonding) are all critical aspects that influence the multitude of phenomena arising from the interplay between electrons and atomic lattices in materials. Electron/lattice interactions are also key in engineering applications, including adhesion, cohesion~\cite{guin2016atomistically}, debonding~\cite{gall2000atomistic} and fracture~\cite{xu2012coupled} in materials under standard and environmental conditions such as irradiation~\cite{serebrinsky2004quantum}, embrittlement~\cite{barrera2018understanding} and to investigate the impact of defects~\cite{khare2007coupled} on material reliability.  

While a rather comprehensive understanding of electron-phonon coupling effects has been achieved in condensed-matter physics~\cite{giustino, hofmann, pintschovius}, little is known about the interplay between nuclear displacements and electronic charge fluctuations at the scale of engineering materials~\cite{curtin2003atomistic,Hauseux}. Even when it is recognized that quantum-mechanical forces at the atomic scale are crucial to reliably determine the mechanics of materials at the macroscopic scale, quantum physics and continuum mechanics models are often developed independently. Quantum-mechanical methods are typically restricted to the modeling of small and well-ordered systems, whereas mechanical properties in mesoscopic and macroscopic engineering problems are routinely quantified resorting to (semi-)classical pairwise (PW) potentials. 
\begin{figure}
\begin{center}
\includegraphics[scale=0.27]{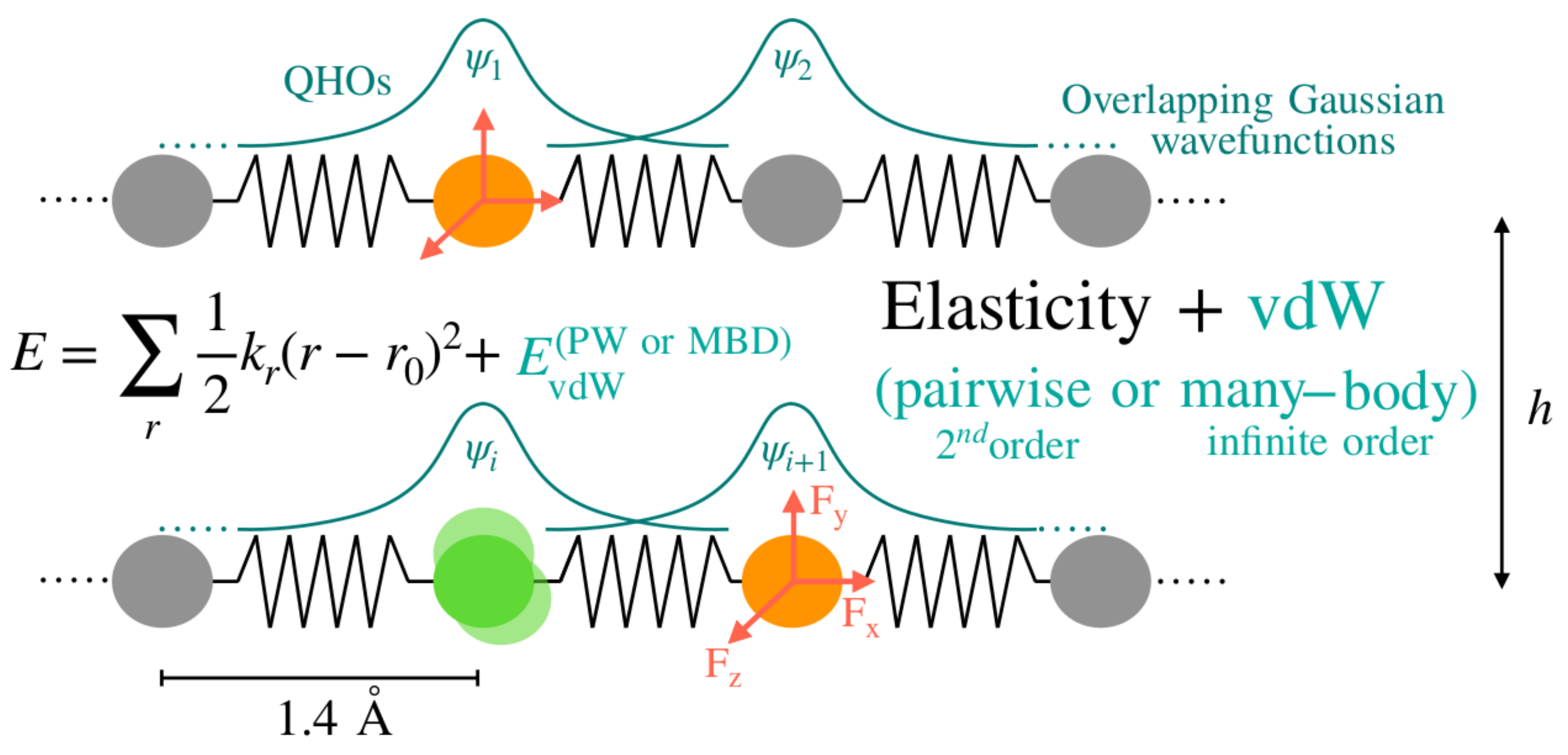}
\caption{
Visual depiction of the geometry of the studied system and the interatomic interaction model. Two carbyne chains with carbon--carbon nearest-neighbor distance of 1.4 {\AA} are separated by a vertical distance $h$. The van der Waals (vdW) interactions between all atoms are treated via either a pairwise or a many-body model based on overlapping quantum harmonic oscillators interacting through a dipolar potential (see text for details). The Hessian matrix elements are computed analytically and correspond to measuring the force response at atom $m$ (yellow atoms) to the displacement of an atom $n$ (green atom). For geometry relaxation and computing adhesive properties, the chemical bonds (local elasticity) are modeled via harmonic springs.}
\label{fig:overview_Hv}
\end{center}
\end{figure}

In this Letter, we study systems of one single chain and two interacting carbyne-like chains (see Fig.~\ref{fig:overview_Hv} for the explanation of the geometry and the atomic interaction model). We focus on the analysis of the atomistic Hessian matrix, which measures the force response on an atom resulting from a perturbation of itself (diagonal terms) or a different atom (off-diagonal terms). The Hessian matrix gives access to computing many response properties, such as phonon eigenvalues and eigenvectors and free energies. We find that a quantum many-body description of vdW interactions yields atomic force response magnitudes that exceed the expected pairwise decay by 3-5 orders of magnitude for a wide range of separations between the perturbed and the observed atom. This is in contrast to ratios between many-body and pairwise vdW interaction energies for fixed structures, which rarely exceed an order of magnitude~\cite{Ambrosetti1171}. Linear chains are analyzed in this study to facilitate physical insight into complex many-body mechanisms. However, key features of intricate coupling between atomic displacements and plasmon-like degrees of freedom are expected to persist in a wide range of low-dimensional nanostructures. Our theoretical study is motivated by experimental evidence of intriguing mechanical properties of systems at the nanometer and micrometer scales, due to their enhanced surface-to-volume ratio, high flexibility, and peculiar response properties~\cite{novoselov2005two,zhangyb2005experiment,balandin2008superior,lee2008measurement,TMC-science-2018,jpcl2019,na2015selective,na2016clean,xin2017adhesion,na2016ultra,loskill2012adhesion}. The explanation of many of these experimental findings is expected to require accounting for many-body vdW correlation terms~\cite{dobson2006prl,Ambrosetti1171,ambrosetti2017prb, podgornik}, and explicit inclusion of the complex geometrical distortions arising at the nanoscale~\cite{Hauseux}. However, fundamental understanding of the intricate coupling between electronic fluctuations and collective atomic displacements has not yet been achieved.

We start by computing and analyzing the Hessian force response matrix corresponding to pairwise (PW) and many-body vdW interactions. In what follows, we will use upper indices to refer to Cartesian $\{x,y,z\}$ components, lowercase lower indices to refer to atoms, and uppercase lower indices to indicate normal modes. The $a$-th Cartesian component of the PW force acting on the $i$-th atom is computed as
\begin{equation}
F_i^{{\mathrm{PW}},a} = 
\partial_{r_i^a}\left(\sum_{j\neq i}^N f_{\rm damp} \cdot \dfrac{C_{6,ij}}{r_{ij}^{6}} \right) \,,
\label{eq:pw_maintext}
\end{equation}
where $r_{ij}=|\mathbf{r}_i-\mathbf{r}_j|$ is the distance between atoms $i$ and $j$ ($r^a_i$ being the $a$-th cartesian component of $\mathbf{r}_i$), $C_{6,ij}$ is the corresponding vdW coefficient~\cite{PhysRevLett102073005}, and ${f_{\rm damp}}$ is a short-range damping function. 
Within the many-body dispersion (MBD) method, instead, the electronic dipole response is mapped into a set of atom-centered quantum harmonic oscillators (QHOs), coupled by a dipolar potential~\cite{proof2013jcp}.
The interaction tensor $\boldsymbol{C}_{ij}^{\mathrm{MBD}, ab}$ is composed by $N^2$ ($3 \times 3$) blocks that account for the coupling between each pair of atoms $i$ and $j$ (with cartesian indices $a$, $b$):
\begin{equation}
    \boldsymbol{C}_{ij}^{\mathrm{MBD},ab} = \omega_i^2 \delta_{ij} \delta_{ab} + (1-\delta_{ij}) \ \omega_i\omega_j \sqrt{\alpha^0_i\alpha^0_j}\ \boldsymbol{T}_{ij}^{ab}.
    \label{eq:MatrixE_mbd_maintext}
\end{equation}
Here $\alpha^0_i$ and $\omega_i$ are
the static dipole polarizability and characteristic frequency of the $i$-th atom, while $\boldsymbol{T}_{ij}^{ab}$ is the dipole-dipole tensor for two overlapping QHOs modeling atoms $i$ and $j$. 
Diagonalization of the interaction tensor $ \boldsymbol{S}^T \boldsymbol{C}^{\mathrm{MBD}} \boldsymbol{S} =\boldsymbol{\Lambda}$ (where $\boldsymbol{\Lambda}_{IJ}^{ab}=\delta_{IJ}\delta_{ab}(\tilde{\omega}_I^a)^2$) yields the $3N$ collective dipole-oscillation modes of the system (via the transformation matrix $\boldsymbol{S}$) and the corresponding interacting frequencies $\tilde{\omega}_I^a$.

The MBD interaction energy is computed as the QHO ground-state energy shift caused by the dipolar interaction
\begin{equation}
E_{\text{c,MBD}} = \hbar \left(\sum_{I=1}^{N}\sum_{a=\{x,y,z\}} \tilde{\omega}_I^a/2 - \sum_{j=1}^{N} 3\omega_j/2\right)\,.
\label{eq:mbd_energy_QHO_maintext}
\end{equation}
The MBD atomic force acting on atom $i$, $F^{{\rm MBD},a}_i=- 
\partial_{r_i^a}E_{\text{c,MBD}}$, can be written as
\begin{equation}
F^{{\rm MBD},a}_i=-\frac{1}{4} 
\left[ (\boldsymbol{\Lambda}_{JJ}^{bb})^{-1/2}\, \boldsymbol{S}_{Jl}^{T\,bc} \left(\partial{r_i^a}\boldsymbol{C}^{\mathrm{MBD}}\right)_{lm}^{cd}\boldsymbol{S}_{mJ}^{db}\right]\,,\\
\label{eq:mbd_force_QHO_maintext}
\end{equation}
where repeated indices are contracted (the same convention will be adopted hereafter).
By definition, the MBD dipole-fluctuation modes have collective character, and this is reflected in the above force expression:  Eq.~\eqref{eq:mbd_force_QHO_maintext} involves both dependence on {\it local} atomic indices ($l,m$) and {\it collective} degrees of freedom (MBD mode index $J$).
The non-locality of plasmon-like MBD modes implies that a change in the position (as well as mass and/or oscillator frequency) of a single atom could produce force response throughout the entire system. 

\begin{figure*}
\hspace*{1cm}
\subfigure{\includegraphics[scale=0.48]{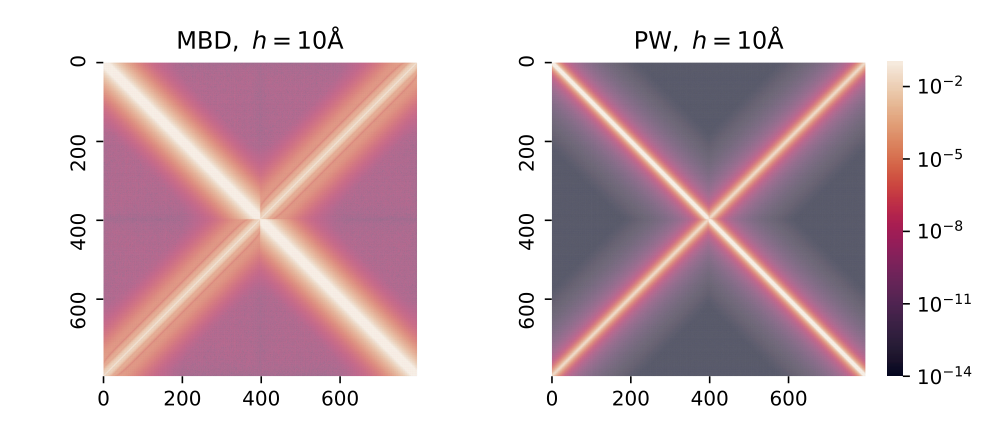} \label{fig:heatmap_PWvdMBD}} 
\put(-250,10){\color{black}(b)}
\put(-460,10){\color{black}(a)}
\put(-389,4){\color{black}Atom index}
\put(-169,4){\color{black}Atom index}
\put(-60,1){\large\color{black}$\mathrm{meV/}$\text{\AA}\textsuperscript{2}}
\\
\subfigure{\includegraphics[scale=0.6]{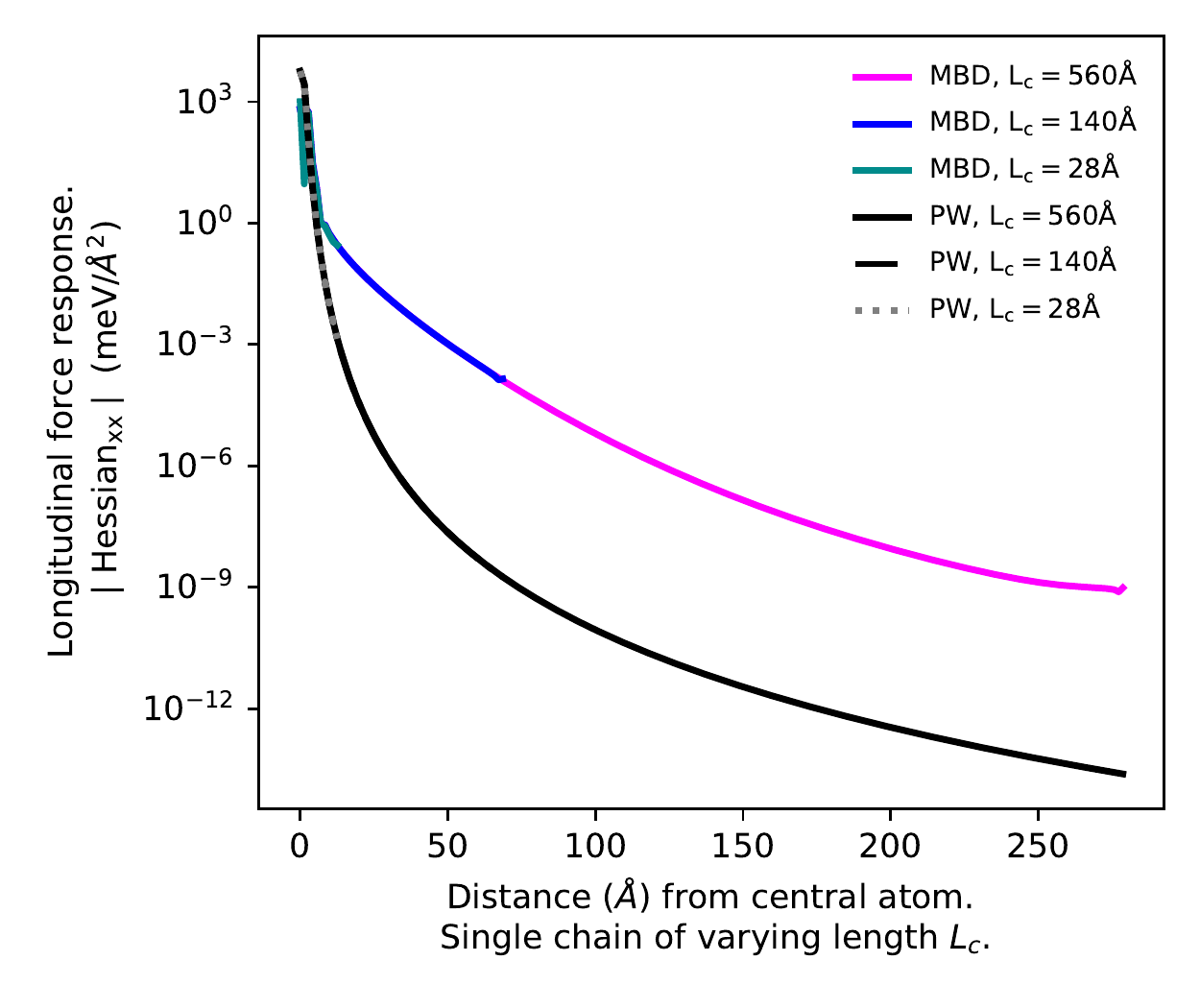} \label{fig:1d_1c_x_x}} 
\subfigure{\includegraphics[scale=0.6]{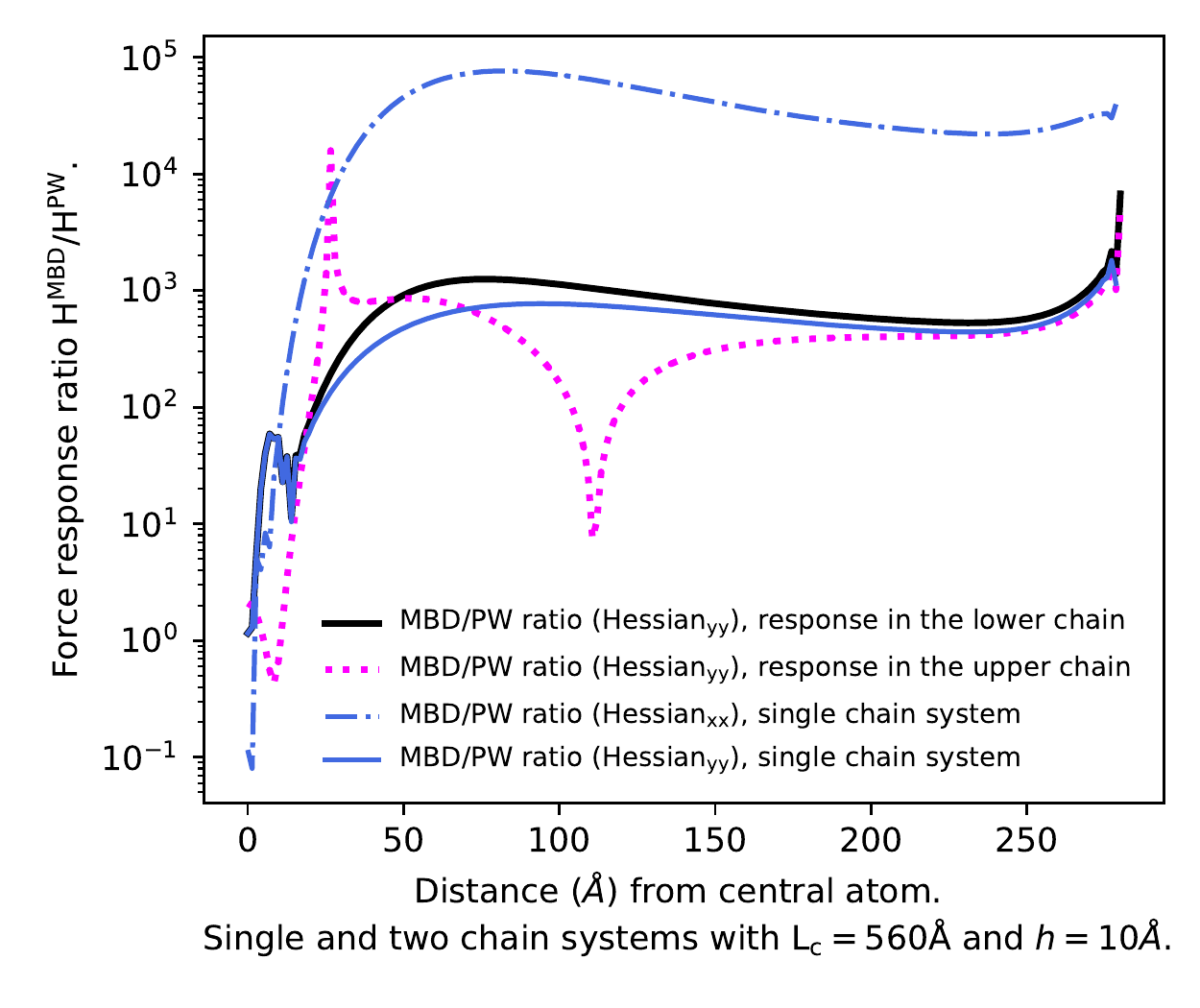} \label{fig:1d_yy_ratio}}
\put(-204,20){\color{black}(d)}
\put(-420,20){\color{black}(c)}
\caption{Heat maps of the condensed Hessian matrix $\boldsymbol{H}^{*}_{ij} = \sqrt{{(\boldsymbol{H}_{ij}^{{xx}}})^2 + ({\boldsymbol{H}_{ij}^{{yy}}})^2 + ({\boldsymbol{H}_{ij}^{{zz}}})^2}$ for (a) MBD and (b) PW vdW interactions.
The Hessian elements measure how much the force at site $j$ changes when an atom at site $i$ is displaced.
Plot axes in (a,b) correspond to $i$ and $j$ atomic sites, respectively. A system of two parallel carbyne chains is considered (with 400 atoms per chain). Here $i,j=(1,400)$ indicate atoms in the first chain, while $i,j=(401,800)$ correspond to the second chain in reversed order. Hence, the main diagonal corresponds to force responses on the same chain, while the minor diagonal corresponds to the chain vertically separated by $h=10$ {\AA}. In panels (c) and (d), one-dimensional Hessian projections are shown upon longitudinal displacement of the central atom in one of the chains. Systems composed of one or two chains with different lengths and different interchain distances were considered at MBD and PW levels (only the case with $h=10$ {\AA} is shown explicitly).}
\label{fig:hessian_1d}
\end{figure*}
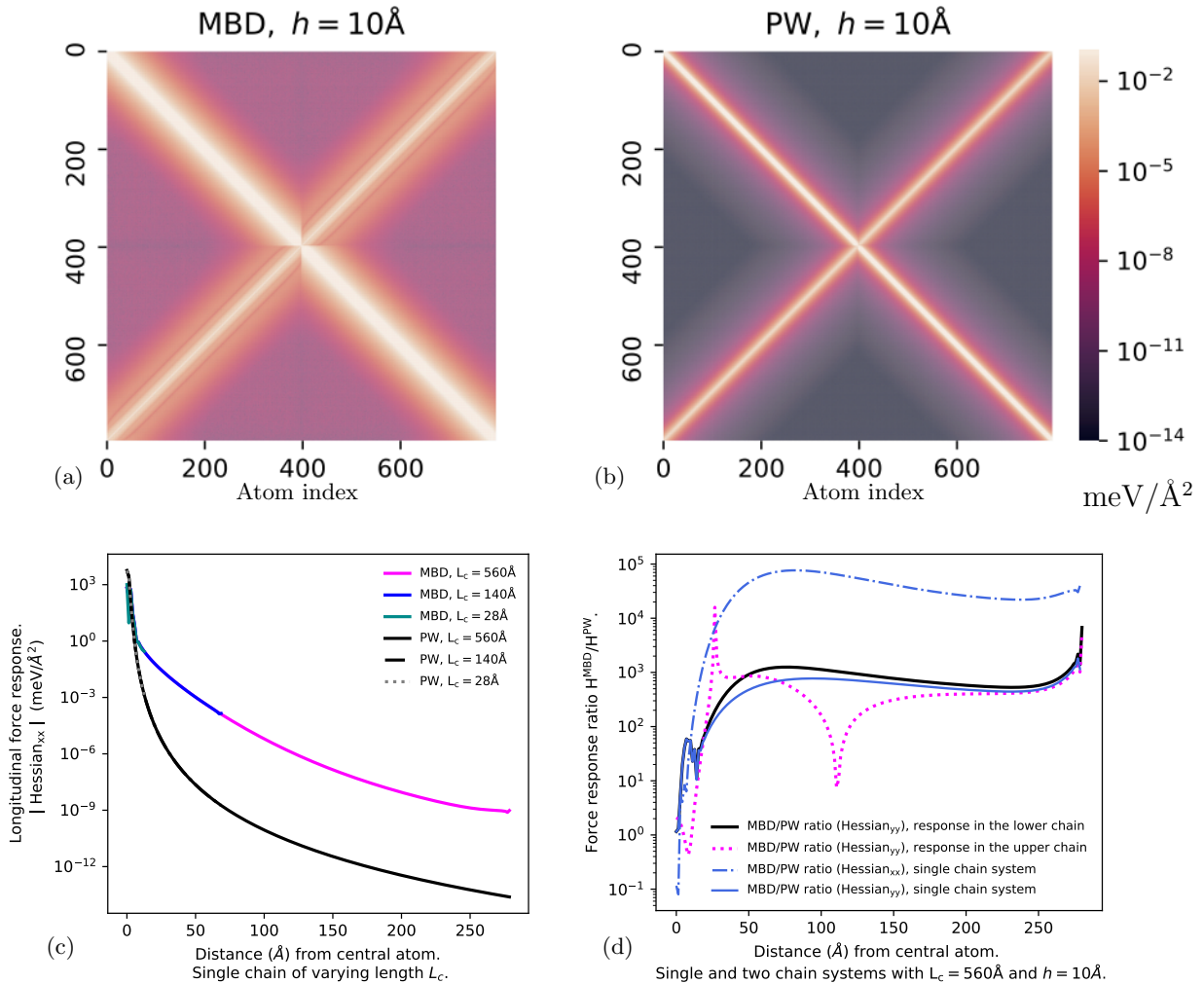

To assess the relevance of the many-body mechanical response, we analyze the vdW Hessian tensor, defined as 
$\boldsymbol{H}^{{\rm MBD}, ab}_{ij}=-\partial{r_i^a}F^{{\rm MBD},b}_j$ (with analogous expression for PW). This tensor quantifies the vdW force response on atom $i$, to an infinitesimal displacement of the atom $j$. The Hessian was computed analytically for both PW and MBD methods. Heat maps of the Hessian matrix for both PW and MBD interactions are plotted in Fig.~\ref{fig:hessian_1d}(a,b). A system of two parallel chains is considered with 400 atoms per chain. In the presence of $N$ atoms, the Hessian is a $3N\times 3N$ matrix. However, for our purposes it is sufficient to analyze the condensed Hessian $N\times N$ matrix $\boldsymbol{H}^{*}_{ij} = \sqrt{({\boldsymbol{H}_{ij}^{{xx}}})^2 + ({\boldsymbol{H}_{ij}^{{yy}}})^2 + ({\boldsymbol{H}_{ij}^{{zz}}})^2}$. The off-diagonal Hessian components that couple longitudinal and transversal degrees of freedom are 5-6 orders of magnitude smaller and have no impact on the forthcoming analysis.

As expected, the displacement of an atom $j$ has the largest impact on neighboring atoms $i$ (see Fig.~\ref{fig:hessian_1d} (a,b)). Accordingly, largest Hessian elements are found on the main diagonal, measuring the force response at the atomic sites closest to the atomic perturbation within the same chain. The minor diagonal, instead, corresponds to the force response on the atoms belonging to the opposite chain (farther away due to finite interchain separation $h$). Comparison between MBD and PW results indicates a substantial role of many-body effects on the force-response nonlocality. 
Moreover, quasi-vanishing MBD Hessian elements found for long interatomic separations (darker regions in Fig.~\ref{fig:hessian_1d} (a, b)) are about three orders of magnitude larger (10$^{-9}$ meV/{\AA}$^2$) than those found at the PW level (10$^{-12}$ meV/{\AA}$^2$). While both of these values may seem small, the substantial difference between MBD and PW has large implications for dynamics and adhesion as we will show below. We also suggest that long-range force response could play a key role at the macroscopic scale and ambient conditions, where the quantity of atoms approaches Avogadro's number and entropic effects always imply a certain degree of disorder at different interatomic length scales.

We now explicitly analyze the response in the longitudinal and transversal components of the atomic forces along the chain induced by a displacement of the central atom. As seen from Fig.~\ref{fig:hessian_1d}(c,d), the MBD and PW force response acts along the chain in a radically different way. Within MBD, Hessian elements exhibit slower decay, and the MBD/PW ratio is always considerably greater than one, in particular at long range with differences of three to almost five orders of magnitude. 

Within the single chain we observe that, after a steep growth due to slower power law decay of the MBD interaction~\cite{Ambrosetti1171}, the MBD/PW Hessian ratio 
tends to saturate (Fig.~\ref{fig:hessian_1d}(d)) beyond the 70 \AA\, scale. This is valid for both longitudinal and transversal Hessian components and suggests that MBD forces exhibit renormalized PW behavior at very long range.  
The renormalization factor is very large, in fact it is necessary to multiply the PW $C_6$ parameter by 
$\sim 10^3$ and $\sim 10^5$ to effectively reproduce the transversal and longitudinal MBD results, respectively. 
Analogous renormalization effects occur also for interchain Hessian elements, \emph{i.e.} for force perturbation due to atomic displacement in the other chain. In the transversal $yy$ case, force response oscillations emerge along the chain. 
This can be analytically understood at the PW level, where a competition arises between a monotonically decreasing $R^{-8}$ factor (due to the second derivative of $R^{-6}$) and a term (proportional to $(7\Delta x^2 - h^2)/(\Delta x^2 + h^2)$, where $R^2=h^2+\Delta x^2$), which is increasing with $x$ (longitudinal distance). 

\begin{figure}
\begin{center}
\includegraphics[scale=0.34]{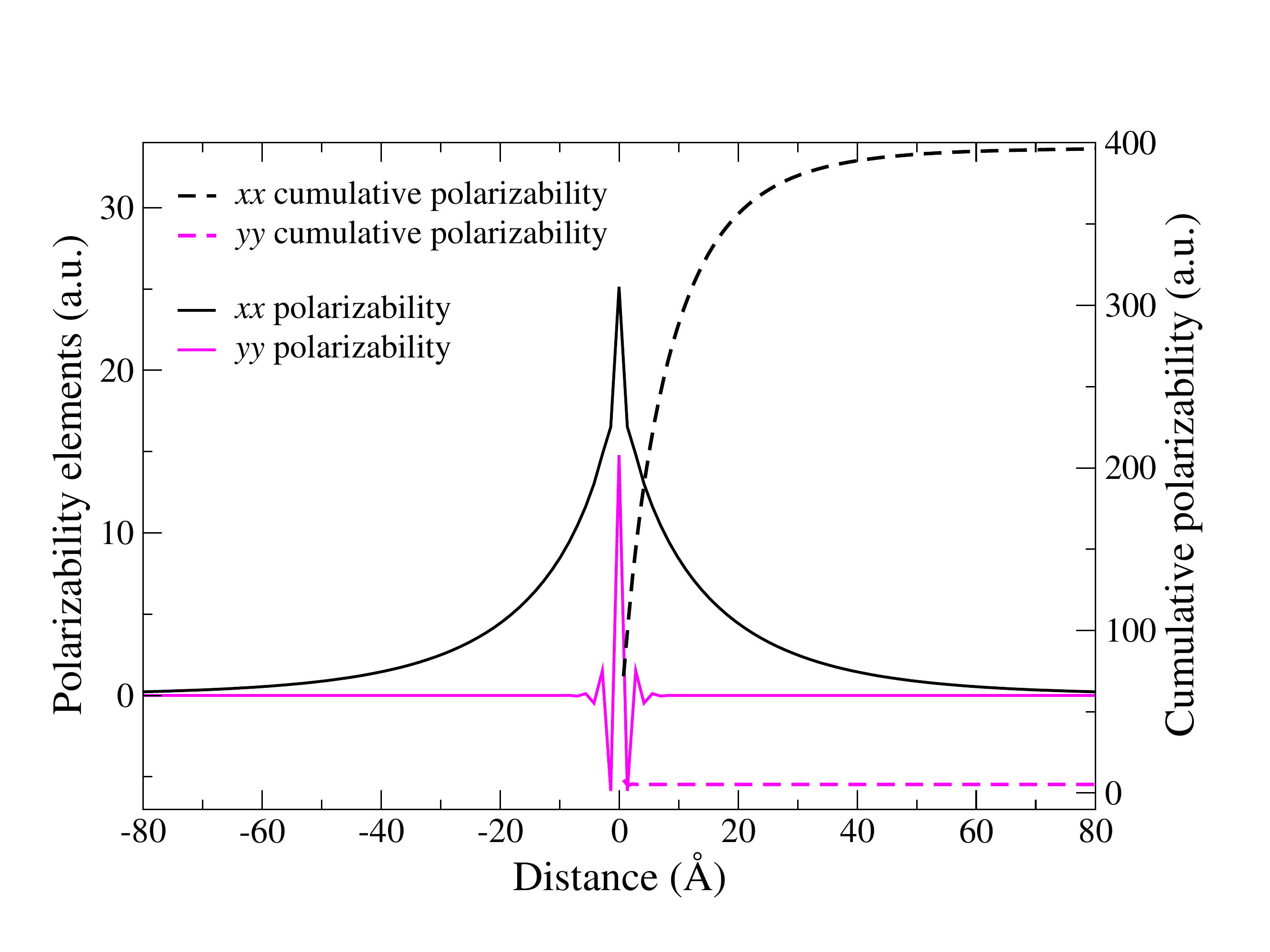}
\caption{Interacting (self-consistently screened) polarizability for a single non-relaxed carbyne chain containing 2001 atoms. Solid lines report polarizability elements $\boldsymbol{\mathcal{A}}_{\overline{i}j}^{aa}$, namely the longitudinal ($a=x$) and transversal ($a=y$) dipole response measured at the atom $j$, located at a given distance from the central atom ($\overline{i}=1001$), due to an external electric field applied in $\overline{i}$. {\it Negative} distances represent the left side of the chain.  Dashed lines correspond to the cumulative dipolar response (sum of all atomic x-x or y-y components, $\sum_{j} \boldsymbol{\mathcal{A}}_{ij}^{aa}$ arising over the entire chain). Here atom $i$ is located at the reported distance from the left chain edge.  
}
\label{fig:pol}
\end{center}
\end{figure}

To rationalize the renormalized pairwise-like behavior of the MBD Hessian elements, we can rewrite the MBD energy in Eq.~(\ref{eq:mbd_energy_QHO_maintext}) using perturbative expansion in terms of the screened polarizability matrix $\boldsymbol{\mathcal{A}}_{ij}^{ab}$ and the dipole coupling tensor $\boldsymbol{T}_{jm}^{bc}$~\cite{proof2013jcp}:
\begin{equation}
E_{\rm c,MBD} \simeq - \int_0^{\infty}\frac{d\omega}{4\pi} \text{Tr}\left[\boldsymbol{\mathcal{A}}_{ij}^{ab}(\mathrm{i}\omega) \boldsymbol{T}_{jm}^{bc} \boldsymbol{\mathcal{A}}_{mn}^{cd}(\mathrm{i}\omega) \boldsymbol{T}_{ni}^{da} \right]\,.
\label{eq:mbd-2ndorder}
\end{equation}
The shown second-order term scales as $\mathcal{C}_{6,jm} r_{jm}^{-6}$ with a screened $\mathcal{C}_{6,jm}$ coefficient and yields the exact MBD interaction energy for widely separated atoms $j$ and $m$, while at shorter distances higher-order terms containing higher powers of $\boldsymbol{\mathcal{A}}$ and $\boldsymbol{T}$ should be considered. We computed the polarizability $\boldsymbol{\mathcal{A}}$ of the single chain using the self-consistent screening (SCS) equation~\cite{MBD-PRL,ambrosetti2014long}. The resulting matrix $\boldsymbol{\mathcal{A}}^{ab}_{ij}$ measures the dipolar response at site $i$ to an electric field applied at $j$. From Fig.~\ref{fig:pol} we observe that the longitudinal ($xx$) polarizability is nonlocal, but has a finite range $\lambda_{\rm C}$ of around $\sim$70 \AA, due to the presence of a finite gap~\cite{Ambrosetti1171} in the dipole excitation spectrum. The renormalization of the single diagonal elements $\boldsymbol{\mathcal{A}}^{xx}_{ii}$ is moderate. However, due to the non-locality of the total polarizability tensor given by off-diagonal elements, about 100 carbon atoms coherently polarize in the presence of a local electric field. The cumulative longitudinal dipole response of these atoms amounts to $\sim 400$ bohr$^{3}$, which corresponds to a polarizability renormalization factor of $\sim 40$ compared to an isolated carbon atom. 
The finite coherence length-scale $\lambda_{\rm C}$ implies that for distances larger than $\lambda_{\rm C}$, different chain fragments will perceive each other as {\it collective polarizability centers}, in analogy to the PW picture. We note that the $C_6$ coefficient for two identical oscillators having static polarizability $\alpha$ and oscillator frequency $\bar{\omega}$ is proportional to $\alpha^2\bar{\omega}$. Hence, by accounting for the polarizability of the coherent chain fragments (characterized by the rescaling factor $\sim$40 and containing 100 atoms), one obtains an effective $C_6$ coefficient renormalization of $\sim$10$^3$, which qualitatively accounts for our numerical observations in Fig.~\ref{fig:hessian_1d}. The even larger renormalization ($\sim 10^5$) found in the $xx$ Hessian elements stems from charge-overlap effects, that are most sensitive to longitudinal geometrical displacements. In contrast, minor renormalization effects are observed in the transverse $yy$ polarizability, while mixed $xy$ terms are essentially vanishing. As shown in Fig.~\ref{fig:pol}, the $yy$ polarizability is far less nonlocal than $xx$, as a consequence of a rapidly damped oscillatory behavior, reminiscent of Friedel's oscillations~\cite{ashcroft}. 

The balance between nonlocal polarization and charge overlap effects depends sensitively on the dimensionality, geometry, and response properties of matter. To show that our findings for chains can be generalized to more complex structures, we carried out Hessian force response calculations for a graphene layer and a carbon nanotube. In both cases, we find renormalization factors $\boldsymbol{H}^{\rm MBD}/\boldsymbol{H}^{\rm PW}$ of $10^3$ to $10^4$ at large interatomic separations, thus confirming the general nature of colossal force response in vdW materials.


\begin{figure}
\begin{center}
\includegraphics[scale=0.7]{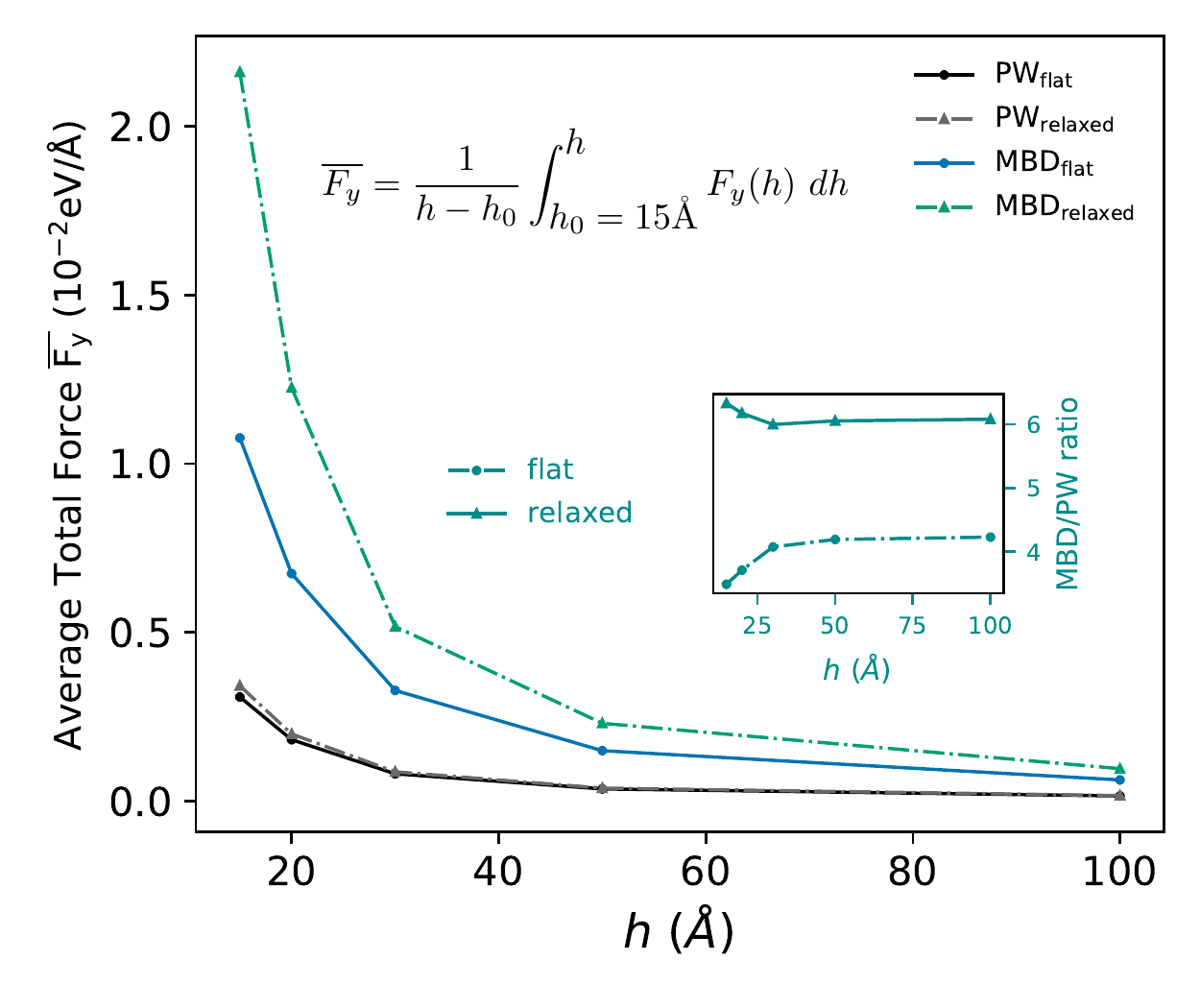}
\caption{Average total vdW transverse force, defined by the given formula, acting on the lower chain as a function of separation distance $h$ in flat and relaxed geometries. MBD/PW average force ratios are also reported for comparison in the inset. The inclusion of MBD interactions consistently leads to stronger adhesive forces compared to the approximate PW method. Upon geometry optimization, PW and MBD forces radically differ, especially at small/medium distances, where the MBD/PW ratio difference is highest. }
\label{fig:F_h_comparison_mt}
\end{center}
\end{figure}


Finally, we demonstrate the importance of the many-body force response for adhesive interactions between two (initially) parallel carbyne chains. We use a hybrid approach introduced in Ref.~\onlinecite{Hauseux} that combines classical elasticity for local chemical bonding and non-local vdW interactions to compute vertical adhesive forces (see Fig.~\ref{fig:overview_Hv} for the depiction of the two-chain model). We compare in Fig.~\ref{fig:F_h_comparison_mt} the total average vdW force acting on the lower chain for two different cases: flat (unrelaxed) configuration and fully relaxed geometry obtained starting from flat parallel configuration constraining edge atoms on their initial positions, at different edge-edge interchain distances $h$. For a set of $h$ values starting from $h_0$ = $15$ \AA, we first compute the total vdW force $F_y$ acting on the lower chain and then compute the average force $\overline{F_y}$ as defined in Fig.~\ref{fig:F_h_comparison_mt}. The inclusion of MBD interactions consistently leads to stronger adhesive forces compared to the approximate PW method. The two chains slightly bend upon relaxation, effectively reducing the interchain separation, and the effect is larger when the MBD method is used. For instance, the average and minimum MBD (PW) interchain separations are 19.75\AA\, (19.96 \AA) and 19.63 \AA\, (19.94 \AA), respectively when $h=20$ \AA. In addition, geometry relaxation using PW vdW forces leads to a negligible change in the adhesion, while relaxing geometries with MBD yields an increase of a factor of two in the adhesive force. We note that much larger adhesive force enhancements have been observed in Ref.~\cite{Hauseux}, however without providing a mechanism for such a finding. The colossal force response in vdW materials arising from long-range electronic correlations provides an explanation for the micrometer-scale adhesive stress observed in experiments~\cite{na2014ultra,na2016ultra,xin2017adhesion} and in calculations that include full geometry relaxation and MBD interactions~\cite{Hauseux}.   

In summary, we have identified a cooperative interplay between electronic correlations and atomic force response in carbyne chains coupled by van der Waals interactions. The nonlocal electronic polarization response and the quantum-mechanical treatment of vdW interactions beyond pairwise approximations is crucial to correctly describe this collective effect. Since many material properties stem from the atomic Hessian matrix, we anticipate that our findings have implications for phonon spectra, free energies, interfacial adhesion, and in general collective dynamics in materials with many interacting atoms.



We are grateful for the support of the Fonds National de la Recherche Luxembourg Grant O17-QCCAAS-11758809. The calculations presented in this paper were carried out using the HPC facilities of the University of Luxembourg. A.A. acknowledges funding from Cassa di Risparmio di Padova e Rovigo (CARIPARO) -- grants EngvdW and Synergy. A.T. acknowledges financial support by the European Research Council (ERC-CoG grant BeStMo). S.P.A.B. thanks the support of the European Union Horizon 2020 research and innovation programme TWINNING Project DRIVEN \url{https://2020driven.uni.lu} under grant agreement No 811099.

P.H. and A.A. contributed equally to this work.

\bibliography{paper.bib}
\bibliographystyle{apsrev4-1}

\end{document}